# False capacitance of supercapacitors

G. A. Ragoisha[1], Y. M. Aniskevich[2]

[1]*Research Institute for Physical Chemical Problems, Belarusian State University, Minsk, 220030, Belarus*

[2]*Chemistry Department, Belarusian State University, Minsk, 220030, Belarus*

**Abstract**. Capacitance measurements from cyclic voltammetry, galvanostatic chronopotentiometry and calculation of capacitance from imaginary part of impedance are widely used in investigations of supercapacitors. The methods assume the supercapacitor is a capacitor, while real objects correspond to different equivalent electric circuits and show various contributions of non-capacitive currents to the current which is used for calculation of capacitance. Specific capacitances which are presented in F $g^{-1}$ units in publications not always refer to electric capacitance. The inadequateness of the capacitance characterization has already resulted in groundless attribution to supercapacitors of various electrochemical systems with electrochemical responses of poorly reversible electrochemical reactions. The number of publications that present false capacitances is terrible and still increases. A widespread neglect of energy dissipation in calculations of specific capacitance leads to further confusion in the characterization of supercapacitors.

## 1 Introduction

Numerous articles have been published lately presenting incredibly high values of specific capacitance $C_s$ (capacitance related to the unit of mass, F $g^{-1}$) which were derived from the current in cyclic voltammetry (CV) and galvanostatic chronopotentiometry, though CV signatures of the electrodes presented typical response of electrochemical reaction, not only the charging current of capacitor. Claims of superior capacitances of those objects as supercapacitors in many cases were also allegedly supported by incongruous calculations of capacitance from imaginary impedance neglecting real part of impedance. The articles increase rapidly in quantity. Table 1 presents typical CV signatures of the objects which were claimed to be excellent supercapacitors, despite the CV presented trivial behavior of faradaic reactions and gave no grounds for capacitance calculation from the electric charge in CV or corresponding galvanostatic chronopotentiometry. The table comprises references to articles bearing "2016" as the year of online or journal issue publication. The time slice corresponds to the period of 2016 by approximately mid-March. Not all latest issues of journals that publish materials about supercapacitors were available to the authors at the time of the manuscript preparation, so the table shows a lower estimate of the damage caused to electrochemical and electrochemistry related journals just in the short period by inadequate publications that disregarded basics of electrochemical methods. The original goal of the review was to draw attention of editors and reviewers to the scientific unfoundedness of those works. By analyzing the articles and reception of similar works in reviews about supercapacitors, we have found that a field with a very specific attitude to the basics of electrochemical methods and interpretation of capacitance in electrochemical systems has already emerged at the interface of electrochemistry and material science. Capacitances about 1000 farad per gram reported for NiO and other materials with typically non-capacitive electrochemical responses are, in fact, a kind of caricature on the methods of specific capacitance measurement which are widely used also for characterization of classical supercapacitors. The dependence on the same improper methods seems to be the reason of a specific attitude (not actively citing, not criticizing) of the "classical" part of supercapacitor community to reports about unusually high capacitances of nickel and cobalt oxides and hydroxides in reviews about supercapacitors.

This article is not the first publication that addresses the problem of false capacitance of supercapacitors. Our analysis of literature was implemented a year after Brousse, Bélanger and Long [1] had pointed to the inappropriateness of nickel and cobalt oxides or hydroxides ascription to pseudocapacitive electrode materials. The article [1] emphasized differences between electrochemical responses of capacitors and materials used in batteries, but inaccurately, in terminology, contrasted capacitive and faradaic nature of electrodes. Fast reversible electrochemical adsorption which corresponds in impedance analysis to Frumkin–Melik-Gaikazyan [2] equivalent electric circuit, with a capacitor in the faradaic branch, can be also called a faradic process, a faradaic process exhibiting pseudocapacitance in terms of Conway [3].

Though most CV signatures in table 1 show no indication of significant pseudocapacitance effect in the faradaic current, and no great capacitance was proved by impedance spectroscopy, the authors of those works used the very fact of the possibility of capacitance relation to faradaic process at unidentified condition as a proof that exactly their objects were supercapacitors, with the consequent attribution of the faradaic current in cyclic voltammetry and galvanostatic chronopotentiometry to the charging current of the assumed capacitance.

The improperly calculated capacitances in articles of this kind are presented usually with incredible accuracy of three to five digits, and the evidently questionable values of specific capacitance often appear in abstract just under title that contains self-advertising phrases ("superior performance", "excellent supercapacitor", etc.), which should have attracted critical attention of reviewers and editors, but the provocation continues with increasing number of publications in various high impact journals. Reviewers, who approved publication of those articles, probably need explanation of what was wrong with the calculation of capacitance from CV and galvanostatic chronopotentiometry and why capacitance should be derived by analysis of impedance spectra, not by direct calculation from imaginary part of impedance; therefore, we will present a summary of basic information related to the capacitance measurement. We will present also typical examples of the reaction of the supercapacitor research community to the critical publication [1], as the reaction to the critique also characterizes the state of research in the field.

## 2 Comments to Table 1

Electrochemical methods were inadequately treated in various respects in literature about supercapacitors. Table 1 presents the latest examples of the most outrageous repudiation of basics of electrochemical methods. In these works, faradaic currents of poorly reversible reactions were attributed to charging currents of capacitance, though the CV signatures prove the objects were not capacitors. Most articles did not present clear arguments why the authors attributed the current entirely to capacitance charging. Surprisingly to electrochemist reader, the most coherent arguments derived the proof of capacitance from a proof of faradaic reaction:

"Nonlinear CP [chronopotentiometry] plots reveals typical Faradaic redox reactions occurring at the electrode/electrolyte sur-/interfaces, further verifying its pseudo-capacitive nature, which is in good line with the CV analysis... Encouragingly, the unique electrode with high loading of 7 mg $cm^{-2}$ delivers superior pseudo-capacitance of ~677, ~631, ~594, ~522 and ~435 F g−1 at the large current densities of 4, 5, 6, 8 and 10 A $g^{-1}$, respectively". This eccentric view on the basics of electrochemical methods is disseminated by Scientific Reports at nature.com [4]. The presence of faradaic reaction was turned by the authors of [4] into a proof of entirely capacitive electrochemical response, which further provided thus accurate calculation of capacitance by galvanostatic chronopotentiometry.

Another example of clear presentation of the substitution of the proof of the capacitance by the proof of faradaic reaction [5]: "All these discharge curves (especially ones with low discharge current densities) show obvious plateaus at 0.3-0.4 V, indicating that the capacitance mainly comes from the Faradaic redox reactions, consistent with the CV curves".

Despite the pointed criticism of article [1] on nickel and cobalt oxide/hydroxide supercapacitors, latest publications about capacitive properties of these materials do not object to critique, but just cite the critical article and continue challenging great capacitances from CV and chronopotentiometry. A characteristic example of inadequate reaction to the critique is article [6] in Electrochimica Acta which gives evidence of the authors being aware of the problem: "Such a terminology "pseudocapacitive" electrode was actually controversially discussed for transition metal hydroxides [1]" (here and further we adjust the reference number in citation to our list of references); nevertheless the article presents the CV signature (Table 1, No 37), definitely not a signature of a capacitor, and the authors derive great "capacitances" of $Ni(OH)_2$ "supercapacitor" from those CV. Moreover, the same article presents impedance spectra with equivalent electric circuit that explicitly shows non-capacitive faradaic branch.

All the proofs of non-capacitive electrochemical response, however, were interpreted as proofs of high capacitance of $Ni(OH)_2$.

One more latest example from Electrochimica Acta [7]: "Recently, it was proposed that NiO or $Ni(OH)_2$ should not be classified as pseudocapacitive material [1]". Despite the acknowledgement of reading the article which warned against using exactly this kind of CV (Table 1, No 38 ) in proof of supercapacitive properties of NiO or $Ni(OH)_2$, the authors of [7] derive high specific capacitance (1,140 F $g^{-1}$) of those materials by integration of CV.

Calculation of specific capacitance of cobalt oxide from CV (Table 1, No 50) using the same formula gave "excellent" result in [8] (687.5 F $g^{-1}$ and 555.2 F $g^{-1}$, incredibly accurate and presented for perfect visibility in the abstract!). The warning against cobalt oxide "supercapacitor" [1] published in the same Journal of Electrochemical Society has not prevented this confusion. Interestingly, even "Warburg resistance" derived from impedance spectra was used to support the capacitance calculation from faradaic current. The authors of [8], as many other in the field, just mix pro and con arguments in their specific logics of proof.

A review article in Inorganic Chemistry Frontiers [9] discusses the problem of wrong supercapacitors in greater detail than other publications which address critique of [1]:

"Brousse et al. proposed that the term "pseudocapacitive" must only be used to describe electrode materials such as $MnO_2$ that displays an electrochemical behavior typical of that observed for a capacitive carbon electrode in a mild aqueous electrolyte. It is confusing that $Ni(OH)_2$ or cobalt oxides which are the same term for materials might exhibit high rate capability, but with the electrochemical signature of a "battery" electrode. [1] ". (Conway [3] had much earlier warned against mixing the signatures of supercapacitors and batteries. The article [1] reminded the well-known information to those investigators of supercapacitors, who probably missed reading the Conway's book.)

"There are situations where the electrochemical behavior of a faradaic electrode can appear pseudocapacitive. The electrode material (powder, thin film etc.) reaches a critical size [10] where diffusion occurs through very limited 5 time scales, giving rise to a "capacitive like" behavior. For $Ni(OH)_2$, there are hundreds of papers dealing with the pseudocapacitive nature of cobalt-based oxides or hydroxides, or even nickel–cobalt oxides or hydroxides, which clearly exhibit a non-linear dependence on the charge stored vs. the potential window."

The article [10] cited in [9], in fact, warned against improper attribution of non-capacitive systems to supercapacitor:

"an electrode material or a device with well-separated redox peaks ... should not be considered a supercapacitor".

The rational idea of [9] is that the electrochemical community should respond to the situation and not be inobservant when hundreds of articles treat as superior capacitors the electrodes that show CV signatures of the kind shown in table 1. The increasing number of those articles is a big confusion to the journals, but the number itself makes no scientific argument. Quality in science is much more important than quantity. Reading a single classical book [11] in due time by those numerous authors would have given them much more for understanding electrochemical methods than cross-citing errors of each other with no proof of the claimed great capacitance in the objects.

## 3 Does "specific capacitance" of supercapacitor characterize electric capacitance?

A capacitor is an electronic device that stores energy in an electric field. The quantitative characteristic of a capacitor capacity to store energy in the electric field is the capacitance *C* which is defined as the amount of separated electric charge that can be stored per unit change in potential:

$$C = Q/\Delta E \quad (1)$$

The impedance *Z* of a capacitor is entirely reactive, i.e. consists of merely imaginary impedance *Z*(im). *Z* is inversely proportional to capacitance *C* and circular frequency $\omega$:

$$Z = (j\omega C)^{-1} \quad (2)$$

$$\omega = 2\pi f, \quad (3)$$

where $j$ is imaginary unit, $f$ – ordinary frequency which is measured in hertz. The other units are farad for *C,* coulomb for *Q*, ohm for *Z.* It is significant that the impedance of a capacitor depends on frequency, while *C* is frequency independent and can be derived by analysis of impedance dependence on frequency.

Figure 1 illustrates the error which arises when impedance of a system with faradaic reaction is considered in terms of “frequency dependent capacitance” which is calculated from imaginary impedance at a certain frequency.

The vertical line of triangles in Fig.1 corresponds to impedance spectrum of a capacitor with $C$=1μF. The impedance spectrum of the same capacitor connected in parallel with a series of resistor $R$=100 Ω and Warburg element with Warburg coefficient $A_w$=100 Ω $s^{-0.5}$ is shown by open circles in the same plot and full circles in the insert with amplified imaginary impedance scale. Both elements which were added to the capacitor in the second circuit are dissipative elements. $R$ and $Z_W$ dissipate energy applied from external source when the capacitor is being charged. (Dissipation in terms of circuit analysis neither excludes nor proves the possibility of $R$ and $Z_W$ to represent electrochemical reaction that stores energy in products as a battery.) Thus a capacitor in the second circuit is less efficient as a device which stores energy in an electric field than the same capacitor in the first circuit. In other words, the addition of the faradaic reaction, when it is represented by $R$-$Z_W$, to a double layer capacitance can only spoil the energy storing capacity of the double layer capacitance. Fantastic values of specific capacitance of the order of 1000 F $g^{-1}$ which were presented in articles and reviews [12], were just result of incorrect interpretation of experimental data, as we will show further.

Three methods of specific capacitance $C_s$ measurement were most commonly referred in publications about supercapacitors: the current integration in CV or measurement of the charge in galvanostatic “charge/discharge cycle” (practically the charges of both “charge” and “discharge” stages are usually obtained by galvanostatic chronopotentiometry) with the further division of the charge by the potential range of CV or chronopotentiometry, and also capacitance calculation from imaginary impedance or, sometimes, impedance modulus at a given frequency. Thus obtained capacitances are referred to the electrode mass or surface area.

Irrespective of differences in equations used for $C_s$ calculation from impedance at a single given frequency, they all give incorrect capacitance data when applied to an object which is not a capacitor or a capacitor connected just in series with a resistor. C and R-C are the only two electric circuits in which imaginary impedance is a measure of capacitance. We will show the basic confusion by applying the method described in [13] to impedance spectra presented in Fig.1. Following [13], we assume

$$C_s = [\pi f_l Z(\mathrm{im})_l m]^{-1}, \quad (4)$$

where $f_l$ is the lowest frequency, $Z(\mathrm{im})_l$ is the imaginary impedance at $f_l$ and $m$ is the weight of electrode. The formula for $C_s$ presented in different publications may differ by coefficient [7,14], which is not significant for our illustration, as both versions do not give capacitance for systems that contain dissipative elements (resistor, impedance of diffusion, CPE) other than a series resistor. The ratio of capacitances, instead of specific capacitances $C_s$, can be considered when comparing objects of equal mass. By substituting the imaginary impedance at $f_l$=0.1 Hz of the two objects presented in Fig.1 (the values are indicated by dotted lines) into Eq (4), one can find that $C_s$ of the object with circuit 2 appears to be four orders greater than $C_s$ of a pure capacitor, due to the corresponding differences of imaginary impedances at $f_l$=0.1 Hz (126 Ω and 1.6 $10^6$ Ω). That is how the capacitor appears to be much greater with the addition of the dissipative elements, just due to the wrong assumption of $C_s$ independence on real part of impedance contained in the use of Eq (4).

Both dc methods of $C_s$ measurement would also show the apparent increase in $C_s$, due to additional charge that bypasses capacitance $C$ via $R$-$Z_W$ branch in circuit 2 in potentiodynamic and galvanostatic experiments. Moreover, the dc methods will show the $C_s$ dependence on the scan rate and the galvanostatic current, as the charge that bypasses capacitance $C$ via $R$-$Z_W$ branch depends on the experiment duration. That is how great capacitances appeared in most investigations of NiO and similar materials with CV signatures shown in Table 1.

The use of Eq (4) or improper application of Eq (2) to objects that are not capacitors also produces false dependences of capacitance on frequency which are presented and discussed in many works related to supercapacitors. The pretended frequency dependence of the false capacitance which results from incorrect treatment of dissipative electric circuits in terms of capacitor should be distinguished from dependence on the frequency range of real capacitors – elements of electric circuit that are derived as capacitors by frequency response analysis. We will return to this thesis in Section 4.

The prerequisite of capacitance calculation by current integration with dc methods should be a proof of the object being a capacitor. Farad is a unit of measurement which is applicable only to capacitor. All articles that present properties of objects in F $g^{-1}$ implicitly assume those objects to be capacitors, but the assumption is not always valid. Measuring properties of objects of other types in farad is the physical nonsense of the same kind as e.g. measuring volume in gram. If the object dissipates some energy when it is being charged, a capacitor can still be the most important part of its equivalent electric circuit, but a capacitor cannot represent the whole dissipative circuit, and that is the catch for very many works in the field of supercapacitors. False capacitances of NiO and similar "supercapacitors" have shown what happens when an object which is not capacitor is characterized by methods accepted in the supercapacitor research. The errors originated not in the objects but in the methods of investigation which groundlessly assumed the objects were capacitors. That is probably the reason of the classical part of supercapacitor community protracted tolerance to the extraordinary violation of basics of electrochemical methods by neophytes. However, the fantastic specific capacitances of new supercapacitors are not widely admitted in the field, e.g. the resent extensive review on supercapacitors [15] references nickel oxide just by two works published in 2003 and 2008, as if there were no latest research on NiO supercapacitors.

Fletcher et al [16] suggested the name "fictional capacitance" for the capacitances that are calculated from imaginary impedance when impedance of the object does not fit to a series *RC* circuit. Eq (2) unambiguously relates imaginary impedance to a capacitance when the object is a solitary capacitor or a series *RC* circuit. On the contrary, various attempts to fit other circuits to a series of resistor and capacitor can produce only "fictional capacitance", as the *RC* model is incompatible in principle with an arbitrary circuit.

Further, we will show how the $C_s$ derivation shows its real worth in the measurement of specific capacitance of "true" supercapacitors, i.e. those supercapacitors which can really store and return energy in the way similar to a capacitor and show, at least very approximately, CV signatures of supercapacitors discussed in [1] and [3]. The latest (in the time of the manuscript preparation) issue of Journal of Solid State Electrochemistry has brought a characteristic article [17] for discussion of that aspect of the problem of the false capacitance. Figure 2 shows a collage based on the data from [17]. The CV (insert to Fig.2a) shows a significant contribution of capacitance by presence of both cathodic and anodic currents at same potentials in almost all potential range of the CV (that is what CV signatures in Table 1 miss). The capacitive current was evidently summed up with non-capacitive potential dependent current, which corresponded to the $R_{ct}$-$Z_w$ branch of the equivalent circuit shown in the same figure. The current in the dissipative $R_{ct}$-$Z_w$ branch by no means gives capacitance, but it was accounted as a part of capacitive current. Thus, the "specific capacitance" (166 F $g^{-1}$) calculated in that work by integration of the sum of capacitive and non-capacitive currents could not have physical meaning of mass specific capacitance. A more detailed examination would have also taken into account the fact that the CPE in the main branch of the circuit not only stores energy in electric field but also dissipates energy [18], so that even the current that passed the CPE should be not entirely accounted for the capacitance (the dissipative nature of CPE becomes evident in its equivalent representation by a ladder composed of capacitors and resistors [16]). This example shows why quantities presented in publications about supercapacitors in F $g^{-1}$ units may not necessarily characterize capacitance.

The specific capacitance as a characteristic of supercapacitor is further compromised by the fact that capacitive and non-capacitive relative contributions to the current used for its derivation can be different in different materials. Thus, $C_s$ is not suitable in general even as a relative characteristic for comparing capacitive properties of objects that correspond to different electric circuits. Great values of $C_s$ can result from small contribution of capacitance to predominantly non-capacitive current, while small $C_s$ can result from predominant contribution of capacitance. Though $C_s$ is usually presented in abstracts of articles as main characteristic of supercapacitor, its real worth is limited by generally unspecified origin of the current from which this characteristic was derived.

## 4 Capacitance in the faradaic branch of equivalent circuit

The motif of many erroneous attempts to derive great capacitance from cyclic voltammograms of the kind presented in Table 1 was the incorrect interpretation of Conway's [3] pseudocapacitance as a capacitance which was supposed to originate by mysterious means from Randles type circuit, i.e. the circuit with no capacitance in the faradaic branch. The pseudocapacitance of a faradaic reaction is in fact a real capacitance [2]. Fig. 3 shows, by circuit 1, equivalent electric circuit of a typical electrochemical system which exhibits the so called pseudocapacitance. The capacitance $C_2$ in the faradaic branch (in parallel with the double layer capacitance $C_1$) is the frequency independent parameter, and the impedances of $C_2$ and $C_1$ can be obtained the same way using Eq (2). The name "pseudocapacitance" which many publications use for this kind of capacitance assumes no discrimination of $C_2$ from $C_1$ by the character of frequency response. The other name of $C_2$ more commonly used in electrochemical literature, "adsorption capacitance" [20], relates $C_2$ to a surface limited faradaic reaction in which this capacitance originates.

A fast reversible surface limited faradaic reaction, e.g. lead underpotential deposition (upd) on gold

$$Pb^{2+} + 2e^{-} \rightarrow Pb_{ad} \quad (5)$$

$$Pb_{ad} \rightarrow Pb^{2+} + 2e^{-} \quad (6)$$

gives CV signature with very small separation of the potentials of the cathodic and anodic peaks (dotted line 1 in Fig.3). The highly reversible reaction provides bidirectional charge transfer in the potential range of the overlapping cathodic and anodic currents, when the electrode is biased in both directions in a wide range of bias rates. Therefore, in frequency response analysis, the formation of lead adatom from lead cation, Eq (5), and the reverse reaction, Eq (6), give three circuit elements: charge transfer resistance $R$ which characterizes kinetics of the charge transfer, impedance of diffusion $Z_W$ which characterizes mass transport, and the capacitance $C_2$ which characterizes the joint capacity of reactions (5) and (6) to accept the applied electric charge and release it back into the circuit the same way as real capacitor does. The charge consumed in reaction (5) can be released fast by reaction (6), so that the electrochemical system examination in terms of electric circuit gives capacitance $C_2$ in a wide range of frequencies, despite the system of reactions (5) and (6), not the electric field created between plates of a capacitor, stores the energy of a capacitor $C_2$. "Pseudo-" in the "pseudocapacitance" $C_2$ refers only to the mechanism of energy storage. Physical methods of electric circuit analysis do not discriminate adsorption capacitance from capacitance of an ordinary capacitor in frequency range of circuit 1 applicability. Thus, "pseudocapacitance" is not a mysterious capacitance of dissipative circuit elements, but a real capacitance which can be recognized by impedance spectrum analysis in terms of equivalent circuits.

In practical aspect, it is significant that $C_2$ is accompanied by charge transfer resistance $R$ and impedance of diffusion $Z_W$. Both elements dissipate energy when the capacitance is charged and their effect should be minimized for $C_2$ application in supercapacitor. Contributions of reactions (5) and (6) to $R$ are inversely proportional to their rate constants [18]. The reaction should be perfectly reversible to provide minimal dissipation, as both the forward and back reactions contribute to $R$. The mass transport can be avoided, when both the parent substance and the product are immobilized on the electrode surface. Many upd processes with anion coadsorption give equivalent electric circuits with no significant impedance of diffusion in impedance spectra analysis [21]. The practical criterion of reversible surface reaction capability to provide pseudocapacitance to a supercapacitor has been long known [3]: a mirror-image appearance of the electrode signature in CV. The dotted line 1 in Fig. 3 shows almost mirror appearance in certain ranges of potential, but the CV signature changes drastically with the electrode surface modification by selenium in the amount close to monolayer before Pb upd (solid line 2). The cathodic deposition of lead in the second system still proceeds in the surface limited mode as upd, i.e. above reversible electrode potential of the reaction

$$Pb^{2+} + 2e^{-} \rightarrow Pb_{bulk}; \quad (7)$$

however, the interaction of $Pb_{ad}$ with selenium makes the system of reactions (5) and (6) to become less reversible, so the CV 2 shows no more mirror-image appearance and the capacitance vanishes from the faradaic branch (equivalent electric circuit 2), in accordance with Conway's criterion.

Experimental observations of the transition between circuits 1 and 2 require the use of nonstationary impedance spectroscopy technique, because of surface limited character of the reactions. The stationary state, e.g. at the potential of the anodic peak in curve 2, corresponds to zero current in both directions and the impossibility to probe any of the reactions by small amplitude ac perturbation; therefore the common stationary EIS would provide no information about kinetics [22], though it can distinguish differences in $C_1$ at potentials of the cathodic and anodic peaks. We refer reader to the review [21] and references therein concerning impedance spectra acquisition and analysis in systems with surface limited reactions.

Finally, we have to comment on the concept of "frequency dependent capacitance" which is widely improperly used in publications about supercapacitors. The capacitance is independent on frequency irrespective of the origin; however, $C_2$ is not a capacitance of ideal capacitor, as it depends on the kinetics of electrochemical reactions. At a certain high frequency the electrochemical reactions may fail to provide the bidirectional charge transfer with the rate required for the capacitance. For convenience of the object treatment in similar terms in different ranges of potential, the imperfect capacitance may be in certain cases considered in terms of "frequency dependent capacitance". The concept of "frequency dependent capacitance" should be used with clear understanding that the accuracy of quantities presented in farad for "frequency dependent capacitance" is limited by the accuracy of the "frequency dependent capacitance" correspondence to the frequency independent capacitance. When strong dependence of capacitance on frequency is presented in publications about supercapacitors, this should be interpreted as the indication of inconsistency between the object and capacitor. Correspondingly, the use of farad is inappropriate in the quantitative characterisation of such objects. The replacement of farad by units of electric charge which is observable in many latest publications on supercapacitors makes just the correction in the use of units, but the charges calculated from CV and galvanostatic chronopotentiometry of "frequency dependent capacitance" still do not characterise those objects as capacitors.

## Conclusions

Capacitance measurements from cyclic voltammetry, galvanostatic chronopotentiometry and calculation of capacitance from imaginary part of impedance are widely used in investigations of supercapacitors. The methods assume the supercapacitor is a capacitor, while real objects correspond to different equivalent electric circuits and show various contributions of non-capacitive currents to the current which is used for calculation of capacitance.

As a result of the incorrect assumptions in capacitance measurements, specific capacitances which are presented groundlessly in $F\ g^{-1}$ units in publications about supercapacitors often characterize joint effect of capacitances and dissipative elements of equivalent circuits with unspecified contributions.

The incorrectness of the capacitance characterization has already resulted in groundless attribution to supercapacitors of various electrochemical systems with electrochemical responses of poorly reversible electrochemical reactions. The number of publications that present false capacitances is terrible and still increases.

Besides the incorrect attribution of non-capacitive faradaic responses to effects of pseudocapacitance, the further confusion in the characterization of supercapacitors results from the widespread neglect of real part of impedance of various origin and corresponding energy dissipation that accompanies charge and discharge of capacitances.

Capacitances should be characterized in terms of equivalent circuit analysis by methods developed in electrochemical impedance spectroscopy to overcome the mess of capacitive, non-capacitive faradaic and various dissipative effects which are presently characterized ambiguously in terms of specific capacitance in publications on supercapacitors.

## References

[1] Brousse T, Bélanger D, Long J (2015) J Electrochem Soc 162:A5185–A5189
[2] Frumkin AN, Melik-Gaykazyan VI (1951) Dokl Akad Nauk 5:855
[3] Conway BE (1999) Electrochemical supercapacitors. Scientific fundamentals and technological applications. Kluwer Academic / Plenum Publishers, New York
[4] Hua H et al (2016) Scientific Reports 6:20973
[5] Zhou Q, Cui M, Tao K, Yang Y, Liu X, Kang L (2016) Appl Surf Sci 365 :125–130
[6] Chen JS, Gui Y, Blackwood DJ (2016) Electrochim Acta 188: 863–870
[7] Abbas SA, Jung KD, Electrochim Acta (2016) 193:145–153
[8] Li J, Liu Z, Li L, Zhu C, Hu D (2016) J Electrochem Soc 163:A417–A426
[9] Li B, Zheng M, Xue H, Pang H (2016) Inorg Chem Front 3:175–202
[10] Simon P, Gogotsi Y, Dunn B, Science (2014) 343:1210
[11] Bard AJ, Faulkner LR (2001) Electrochemical Methods: Fundamentals and Applications, 2nd ed, Wiley, New York
[12] Sk MM, Yue CY, Ghosh K, Jena RK (2016) J Power Sources 308:121–140
[13] Farma R, Deraman M, Awitdrus, Talib IA, Omar R, Manjunatha JG, Ishak MM, Basri NH, Dolah BNM (2013) Int J Electrochem Sci 8: 257–273
[14] Yuan C, Li J, Hou L, Lin J, Zhang X, Xiong S (2013) J. Mater. Chem. A **1**:11145-11151
[15] González A, Goikolea E, Barrena JA, Mysyk R (2016) Review on supercapacitors: Technologies and materials. Renewable and Sustainable Energy Reviews 58:1189–1206
[16] Fletcher S, Black VJ, Kirkpatrick I (2014) J Solid State Electrochem 18:1377-1387
[17] Wolfart F, Dubal DP, Vidotti M, Holze R, Gómez-Romero P (2016) J Solid State Electrochem 20:901-910
[18] Lasia A (2014) Electrochemical Impedance Spectroscopy and its Applications. Springer, New York
[19] Bondarenko AS, Ragoisha GA, Osipovich NP, Streltsov EA (2005) Electrochem Commun 7:631–636
[20] Bard AJ, Inzelt G, Scholz F (eds) (2012) Electrochemical dictionary, 2nd ed. Springer, Berlin: page 481
[21] Ragoisha GA (2015) Electroanalysis 27: 855–863
[22] Cesiulis H, Tsyntsaru N, Ramanavicius A, Ragoisha G, The study of thin films by electrochemical impedance spectroscopy. Chapter 1 in: Tiginyanu I et al (eds.) Nanostructures and thin films for multifunctional applications, Springer International Publishing Switzerland, 2016, Section 1.5, pp 23–26

Table 1. These CV signatures were presented to prove capacitive response of electrodes in articles about supercapacitors published in 2016

| | | | | |
|---|---|---|---|---|
| 1 | 2 | 3 | 4 | 5 |
| Sci Rep (2016) 6: 20021<br>DOI: 10.1038/srep20021<br>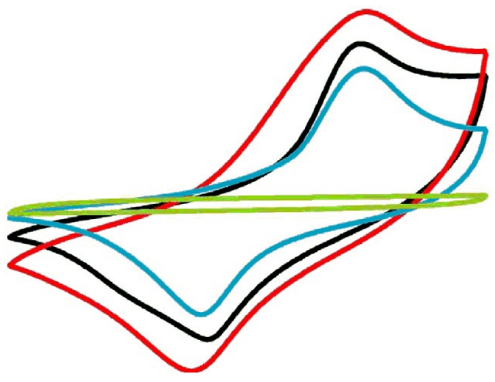 | Sci Rep (2016) 6: 20973<br>DOI: 10.1038/srep20973<br>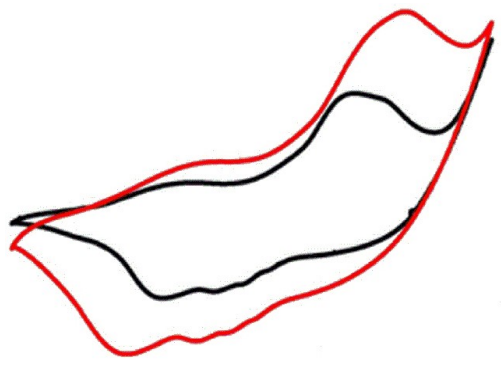 | Sci Rep (2016) 6: 21566<br>DOI: 10.1038/srep21566<br>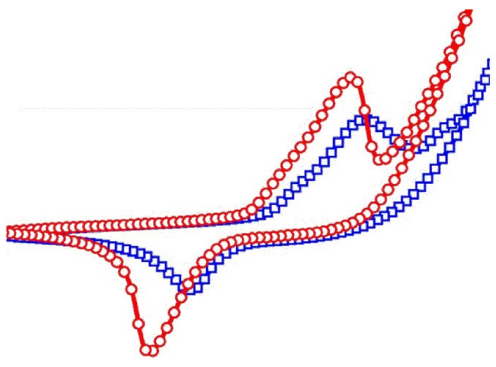 | ChemComm (2016) 52: 4517 DOI: 10.1039/c6cc00215c<br>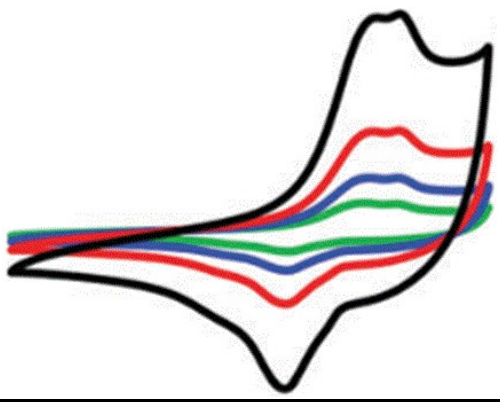 | ChemComm (2016) 52: 2721 (supplement)<br>DOI: 10.1039/C5CC10113A<br>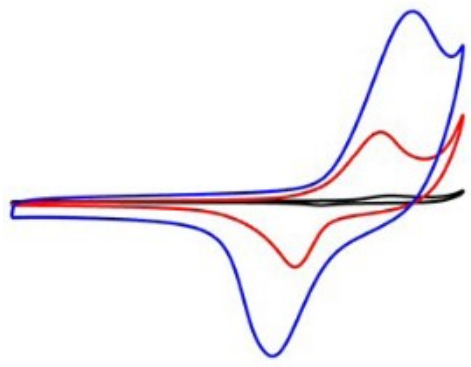 |
| 6 | 7 | 8 | 9 | 10 |
| ChemComm (2016) 52: 3919<br>DOI: 10.1039/C5CC09575A<br>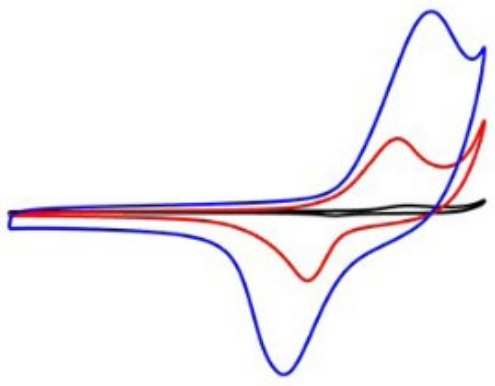 | ChemComm (2016) 52: 2557<br>DOI: 10.1039/C5CC08373G<br>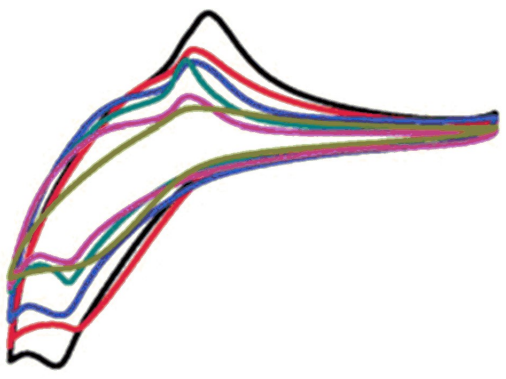 | ChemComm (2016) 52: 1509<br>DOI: 10.1039/C5CC09402J<br>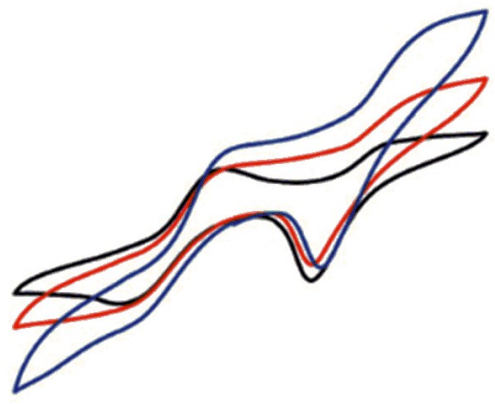 | RSC Adv (2016) 6: 26612<br>DOI: 10.1039/C6RA00426A<br>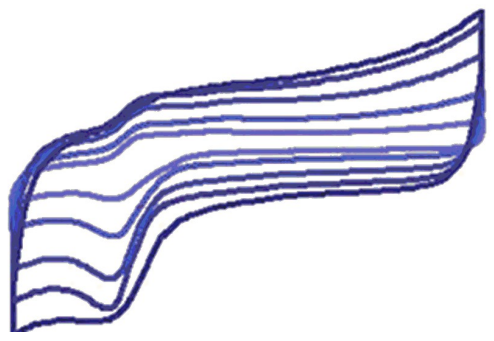 | RSC Adv (2016) 6: 15137<br>DOI: 10.1039/c5ra26391c<br>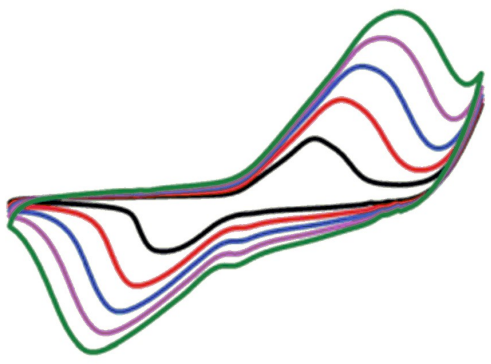 |
| 11 | 12 | 13 | 14 | 15 |
| RSC Adv (2016) 6: 21246<br>DOI: 10.1039/c5ra26946f<br>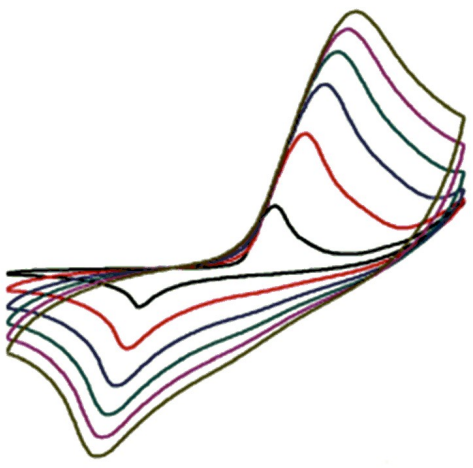 | RSC Adv (2016) 6: 28270<br>DOI: 10.1039/C6RA00004E<br>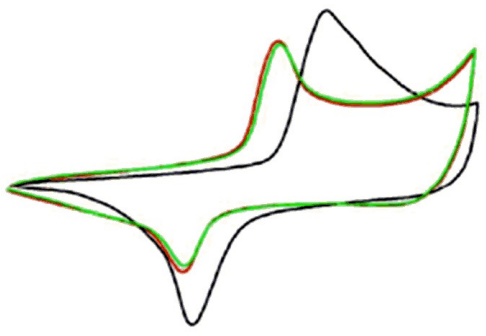 | RSC Adv (2016) 6: 29519<br>DOI: 10.1039/C6RA01474G<br>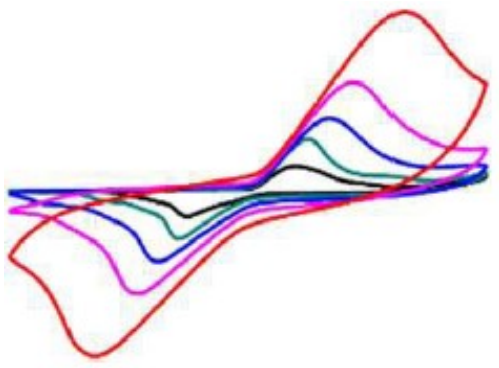 | RSC Adv (2016) 6: 29840<br>DOI: 10.1039/C5RA27375G<br>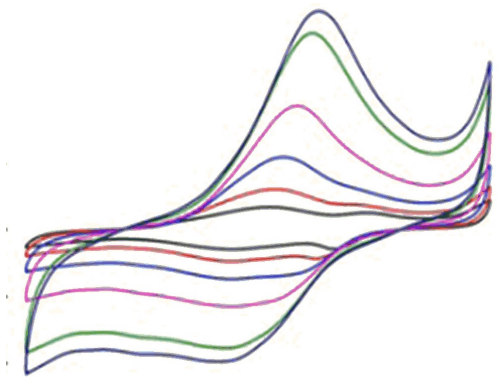 | PhysChemChemPhys (2016) 18: 2718<br>DOI: 10.1039/c5cp06147d<br>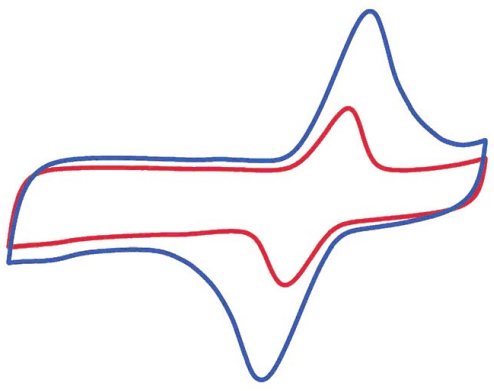 |
| cv | 17 | 18 | 19 | 20 |
| Dalton Trans (2016)<br>DOI: 10.1039/C6DT00002A<br>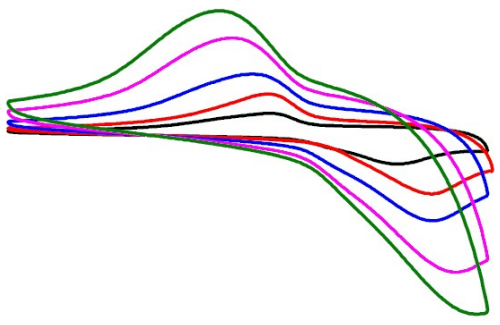 | J Mater Chem A (2016) 4: 4920<br>DOI: 10.1039/C5TA09740A<br>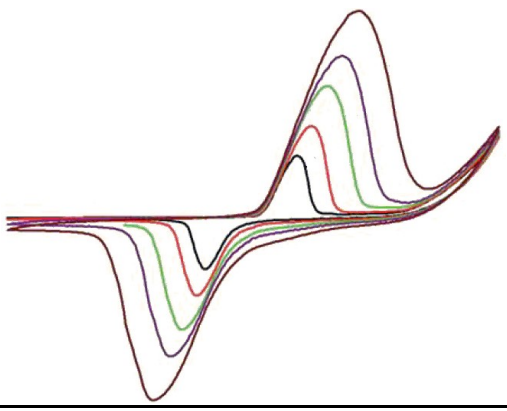 | J Mater Chem A (2016) 4: 2188<br>DOI: 10.1039/C5TA10297A<br>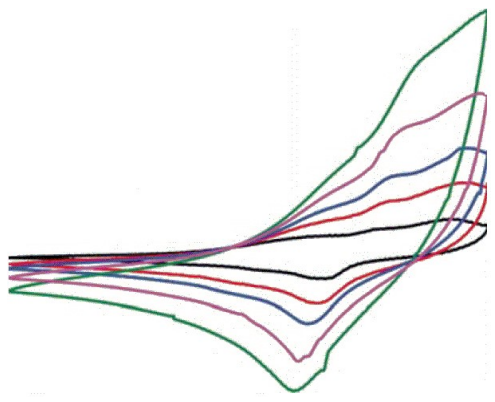 | J Mater Chem A (2016) 4: 4718<br>DOI: 10.1039/C5TA10781D<br>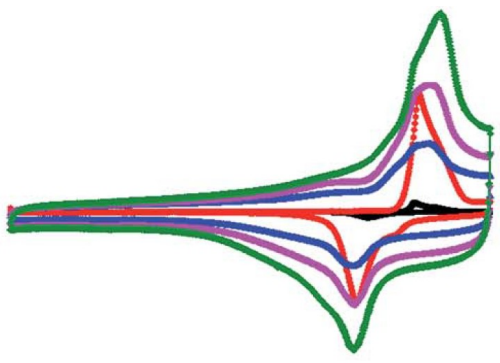 | J Mater Chem A (2016) 4: 3267<br>DOI: 10.1039/C5TA09699E<br>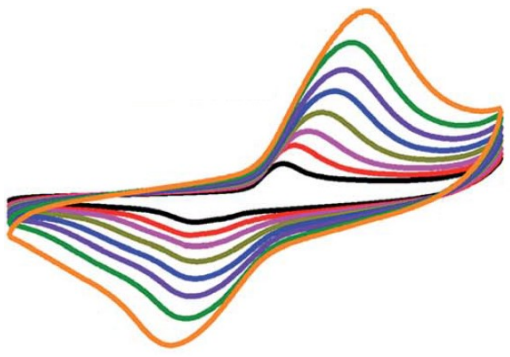 |
| 21 | 22 | 23 | 24 | 25 |
| Appl Surf Sci (2016) 360(B): 534 DOI: 10.1016/ j.apsusc.2015.10.187 | Appl Surf Sci (2016) In press DOI: 10.1016/ j.apsusc.2016.03.106 | Appl Surf Sci (2016) 362: 469 DOI: 10.1016/ j.apsusc.2015.11.194 | Appl Surf Sci (2016) 370: 452 DOI: 10.1016/ j.apsusc.2016.02.147 | Appl Surf Sci (2016) 363: 381 DOI: 10.1016/ j.apsusc.2015.12.039 |
| 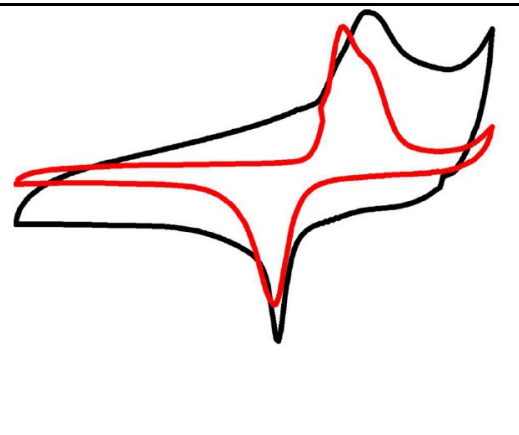 | 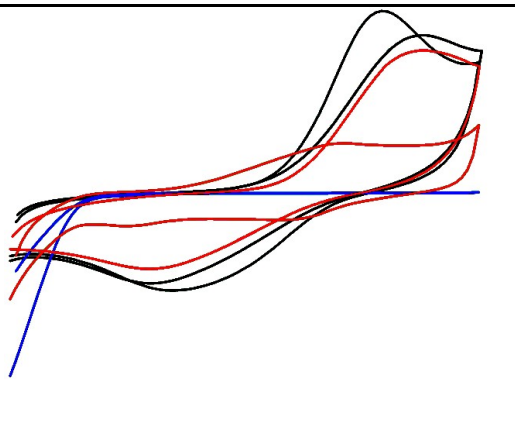 | 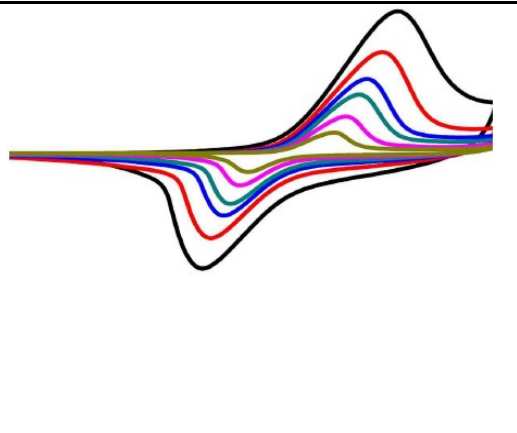 | 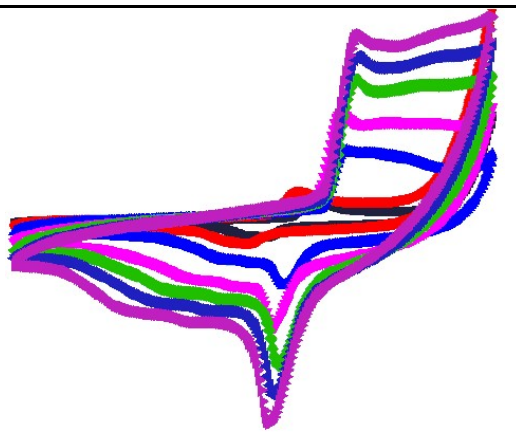 | 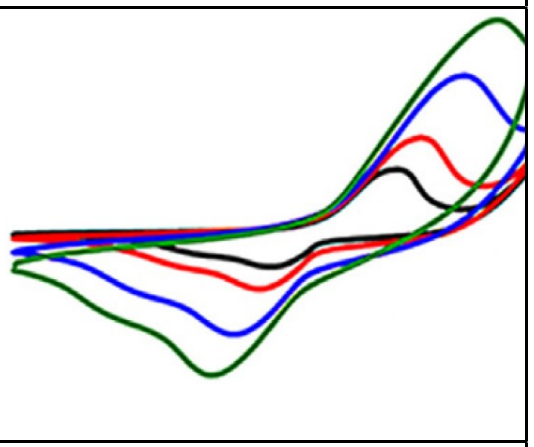 |
| 26 | 27 | 28 | 29 | 30 |
| Appl Surf Sci (2016) 360: 234 DOI: 10.1016/ j.apsusc.2015.11.018 | Appl Surf Sci (2016) 370: 297 DOI: 10.1016/ j.apsusc.2016.02.175 | Appl Surf Sci (2016) 365: 125 DOI: 10.1016/ j.apsusc.2016.01.020 | Appl Surf Sci (2016) 361: 57 DOI: 10.1016/ j.apsusc.2015.11.171 | Mater Des (2016) 97: 407 DOI: 10.1016/ j.matdes.2016.02.114 |
| 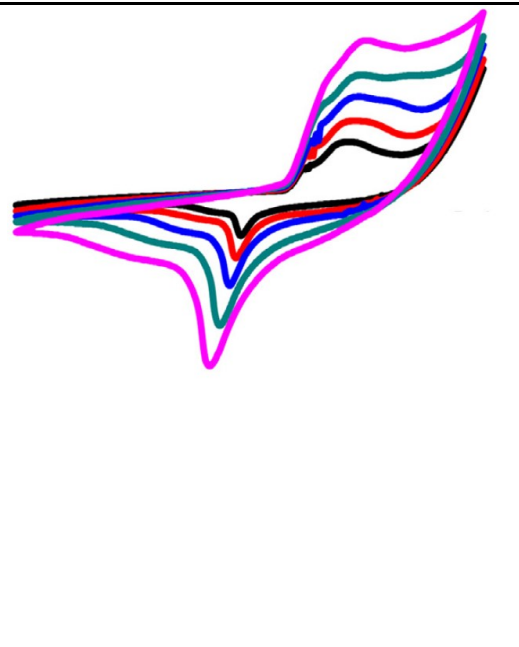 | 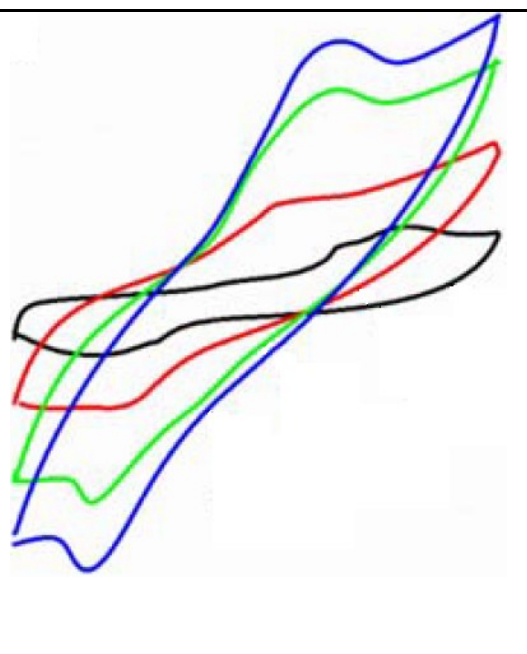 | 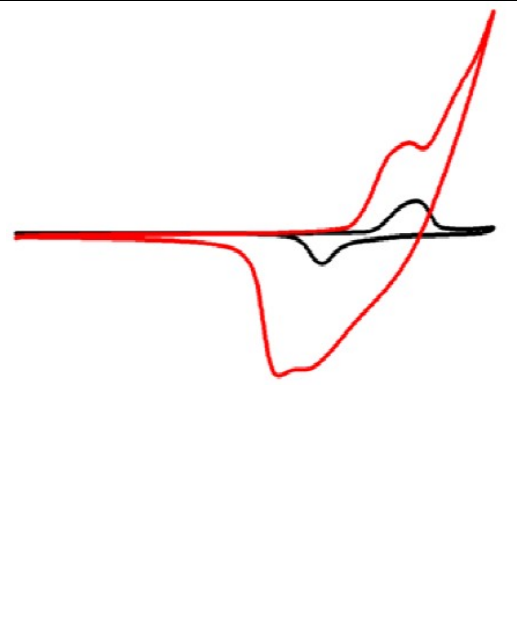 | 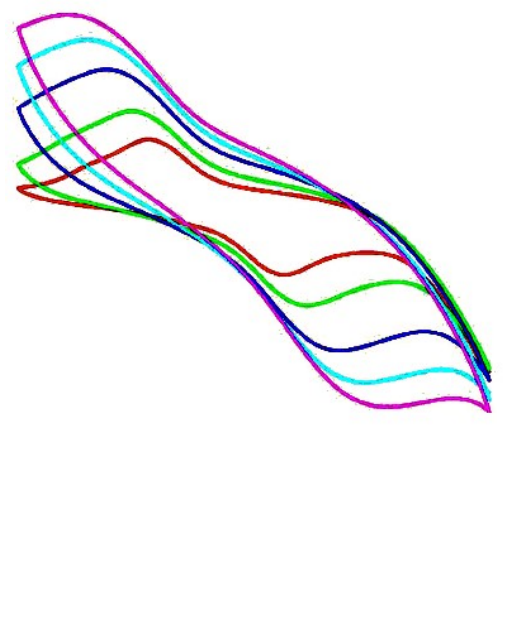 | 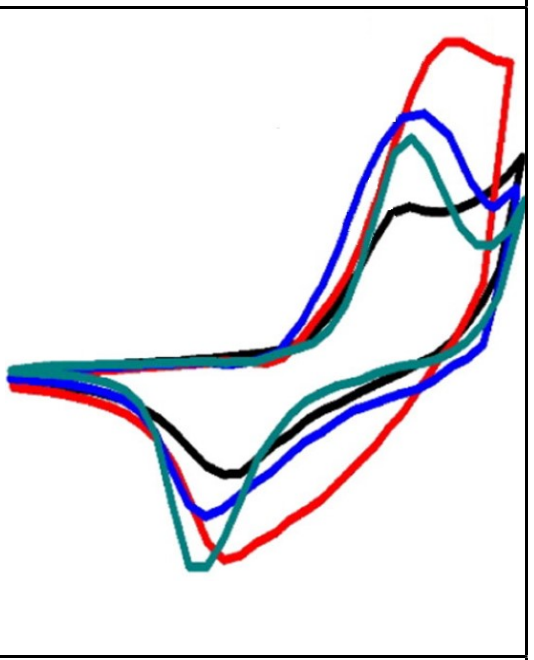 |
| 31 | 32 | 33 | 34 | 35 |
| Electrochim Acta (2016) 191: 275. DOI: 10.1016/j.electacta.2016.01.072 | Electrochim Acta (2016) 192: 205. DOI: 10.1016/j.electacta.2016.01.211 | Electrochim Acta (2016) 191: 364 DOI: 10.1016/j.electacta.2016.01.007 | Electrochim Acta (2016) 191: 444 DOI: 10.1016/j.electacta.2015.12.143 | Electrochim Acta (2016) 193: 104 DOI: 10.1016/j.electacta.2016.02.069 |
| 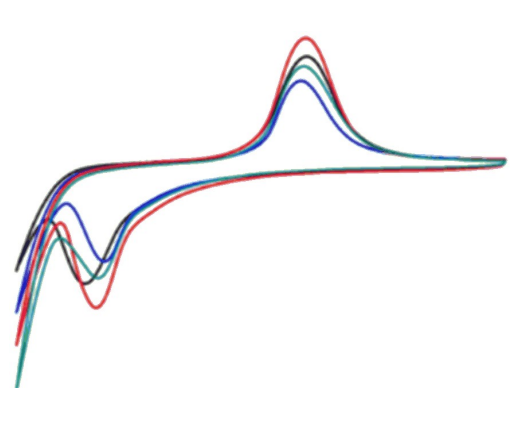 | 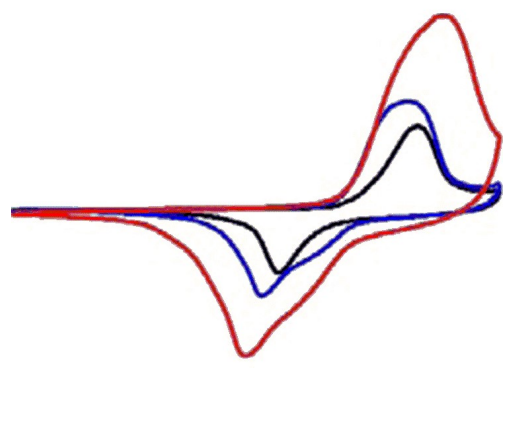 | 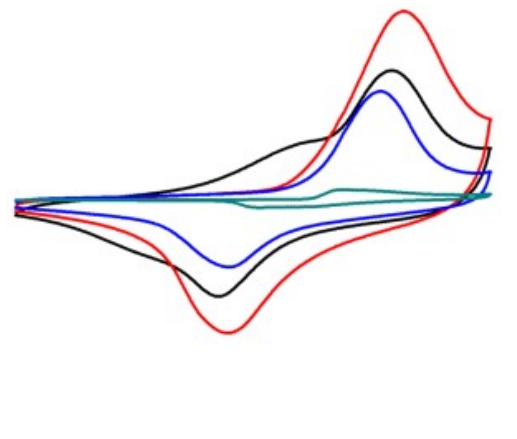 | 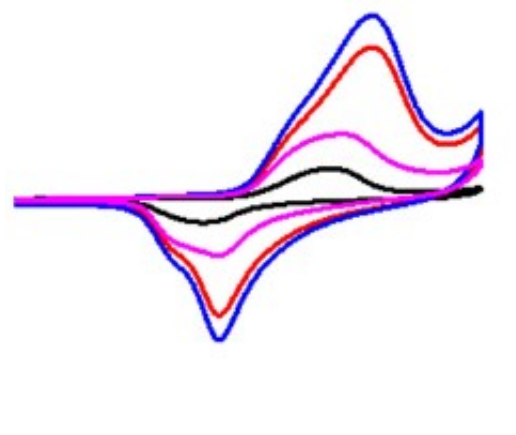 | 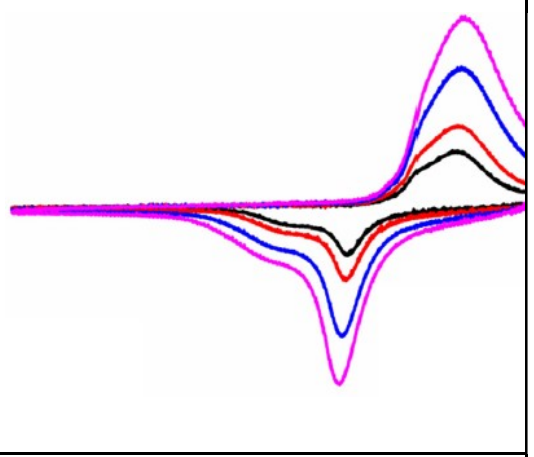 |
| 36 | 37 | 38 | 39 | 40 |
| Electrochim Acta (2016) 188: 13. DOI: 10.1016/ j.electacta.2015.10.165 | Electrochim Acta (2016) 188: 863 DOI: 10.1016/ j.electacta.2015.12.029 | Electrochim Acta (2016) 193: 145 DOI: 10.1016/ j.electacta.2016.02.054 | J Power Sources (2016) DOI: 10.1016/ j.jpowsour.2015.12.029 | J Power Sources (2016) DOI: 10.1016/ j.jpowsour.2015.11.034 |
| 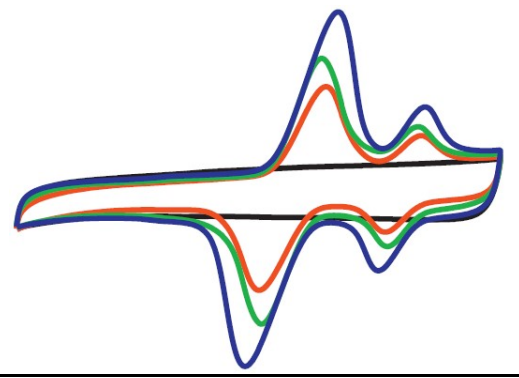 | 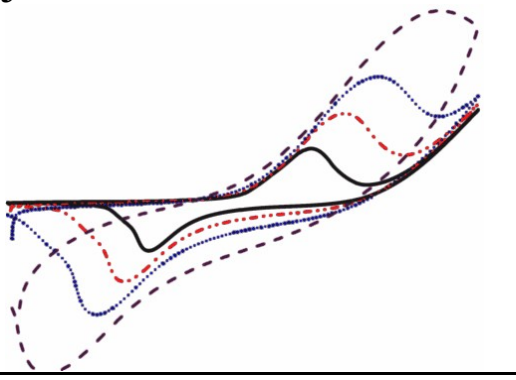 | 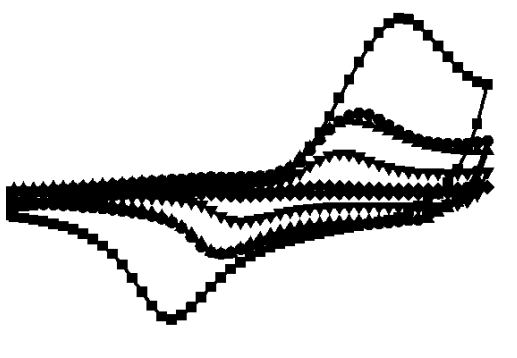 | 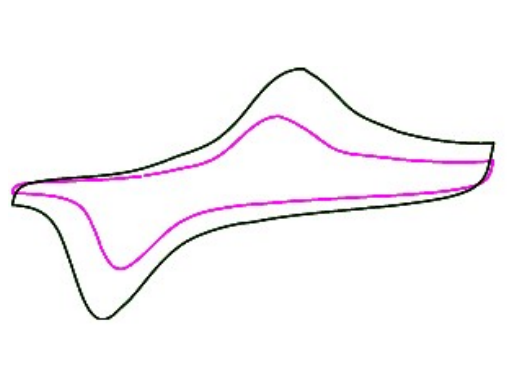 | 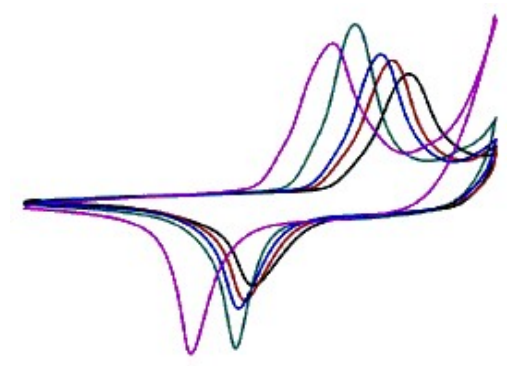 |

| 41 | 42 | 43 | 44 | 4 |
|---|---|---|---|---|
| J Power Sources (2016) 308: 121<br>DOI: 10.1016/ j.jpowsour.2016.01.056<br>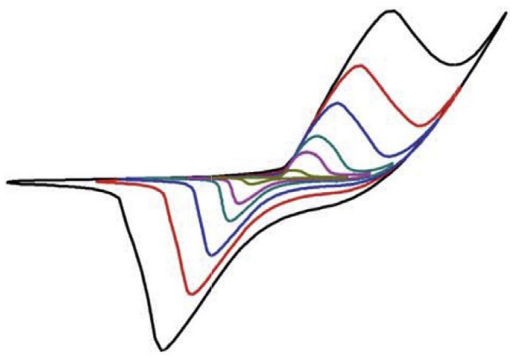 | ACS Appl Mater Interfaces (2016) 8 (5): 3258<br>DOI: 10.1021/acsami.5b11001<br>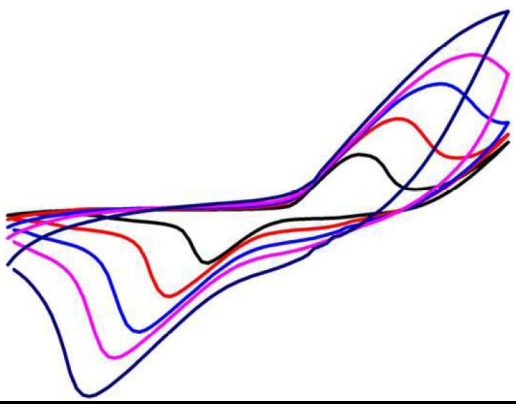 | ACS Appl Mater Interfaces (2016) 8 (7): 4585<br>DOI: 10.1021/acsami.5b10781<br>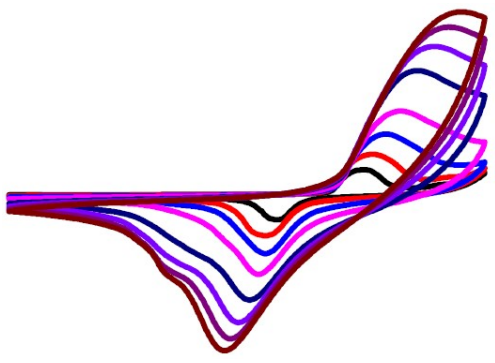 | ACS Appl Mater Interfaces (2016) 8 (4): 2741<br>DOI: 10.1021/acsami.5b11022<br>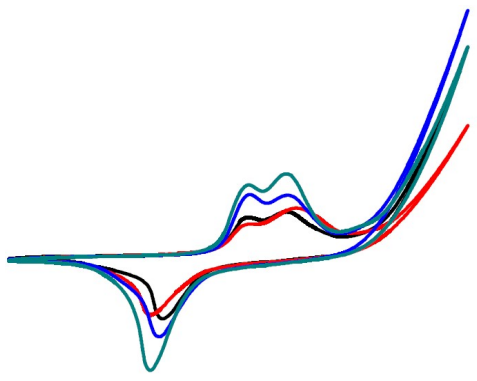 | ACS Appl Mater Interfaces (2016) 8 (9): 6093<br>DOI: 10.1021/ acsami.6b00207<br>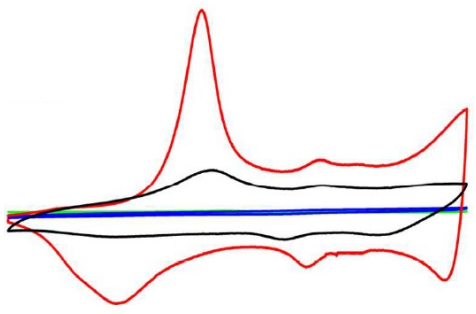 |
| 46 | 47 | 48 | 49 | 50 |
| ACS Appl Mater Interfaces (2016) 8 (1): 780<br>DOI: 10.1021/acsami.5b09997<br>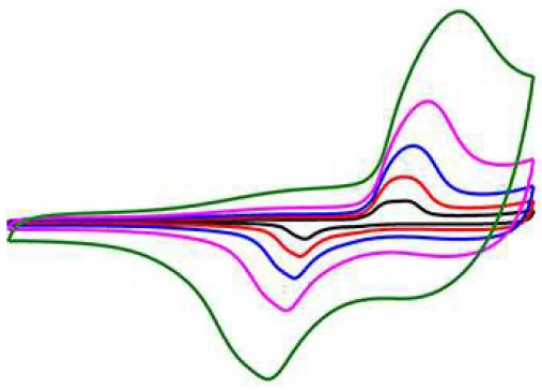 | J Solid State Electrochem (2016) DOI: 10.1007/ s10008-015-3009-2<br>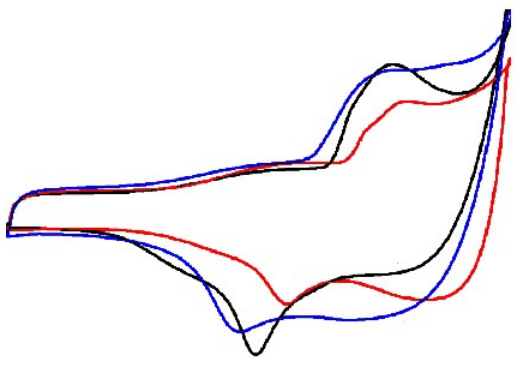 | J Solid State Electrochem (2015) DOI: 10.1007/ s10008-015-3022-5<br>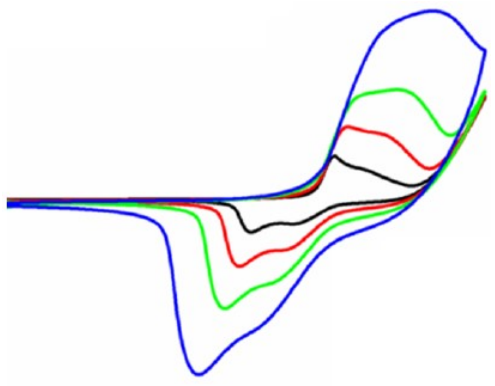 | J Appl Electrochem (2016) 46: 441<br>DOI: 10.1007/s10800-016-0921-9<br>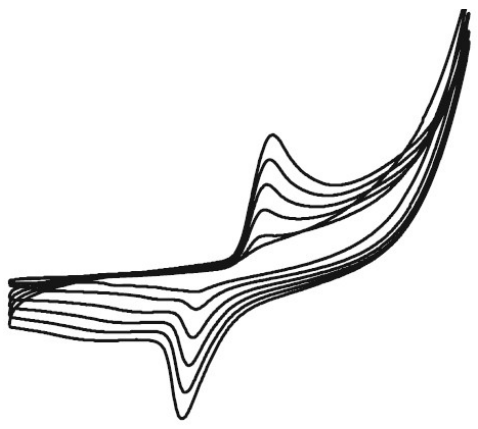 | J Electrochem Soc (2016) 163: A417 DOI: 10.1149/2.0331603jes<br>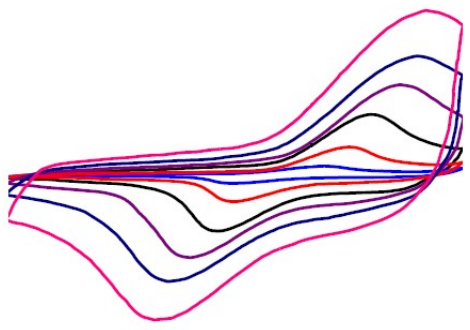 |
| 51 | 52 | 53 | 54 | 55 |
| Int J Electrochem Sci (2016) 11: 1541<br>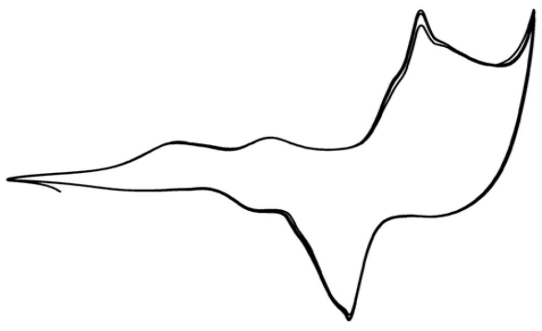 | Chem Eur J (2016)<br>DOI: 10.1002/chem.201504569<br>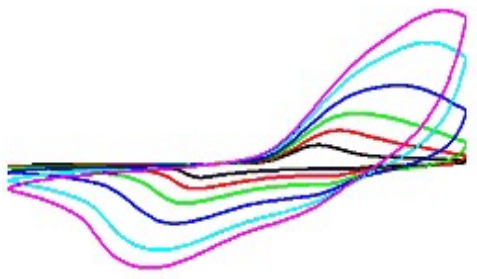 | ChemNanoMat (2016)<br>DOI: 10.1002/cnma. 201600039<br>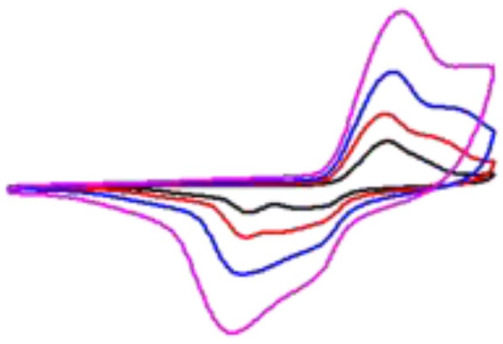 | Mater Res Bull (2016) 76: 229<br>DOI: 10.1016/j.materresbull. 2015.12.023<br>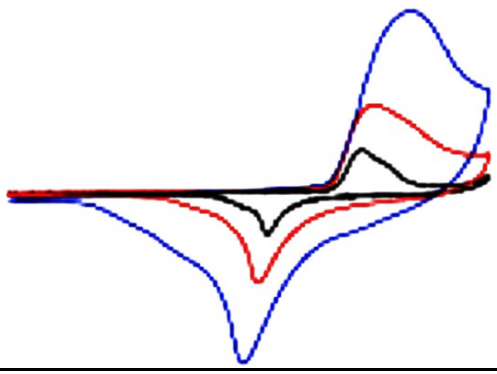 | Mater Lett (2016) 164: 260<br>DOI: 10.1016/ j.matlet.2015.11.011<br>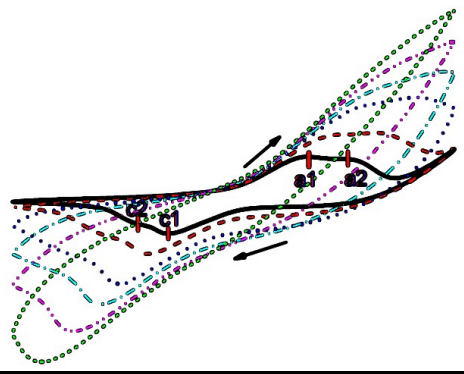 |

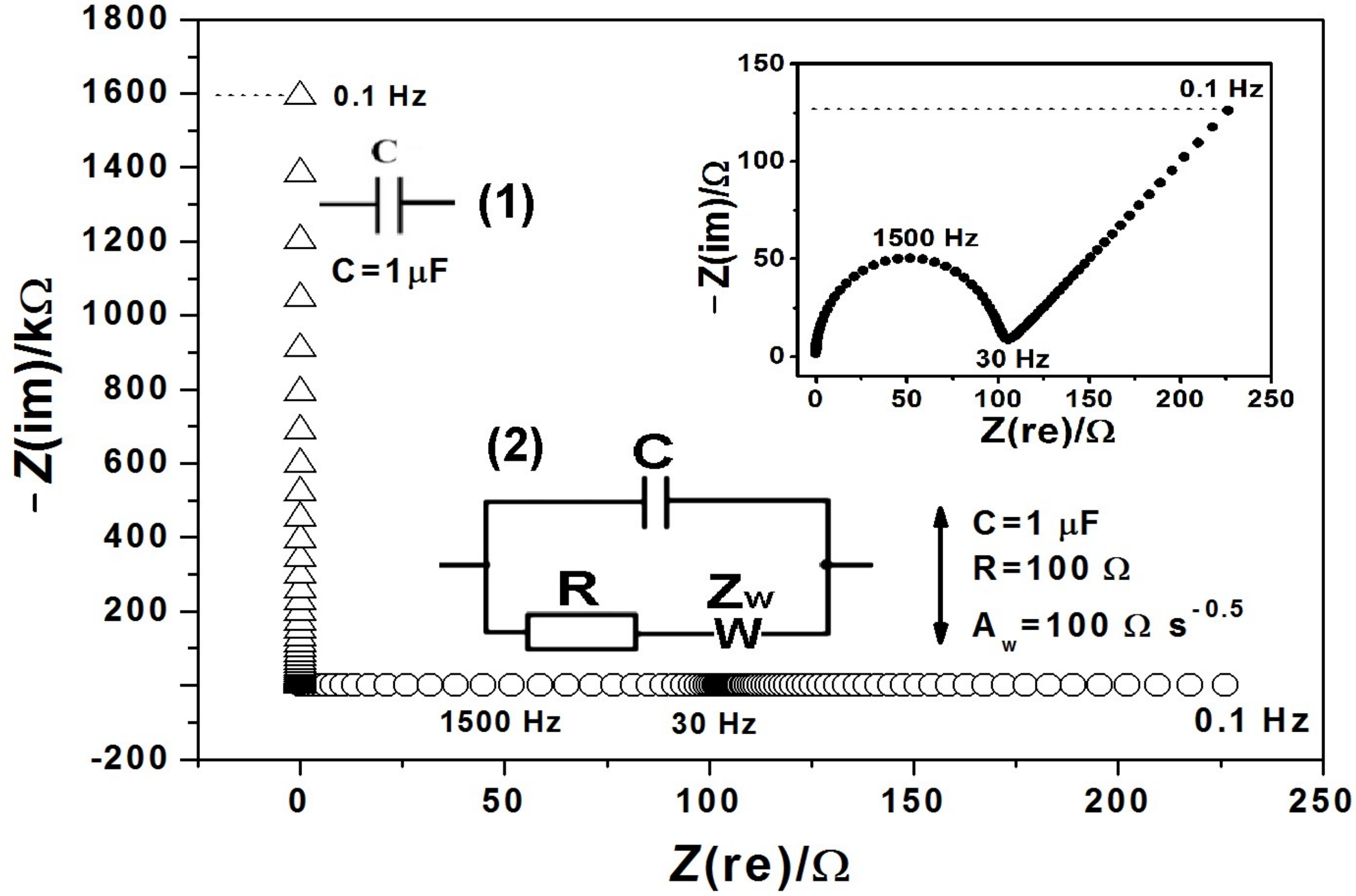

Fig. 1 Illustration of the inappropriateness of capacitances comparison in different types of circuits by imaginary impedance

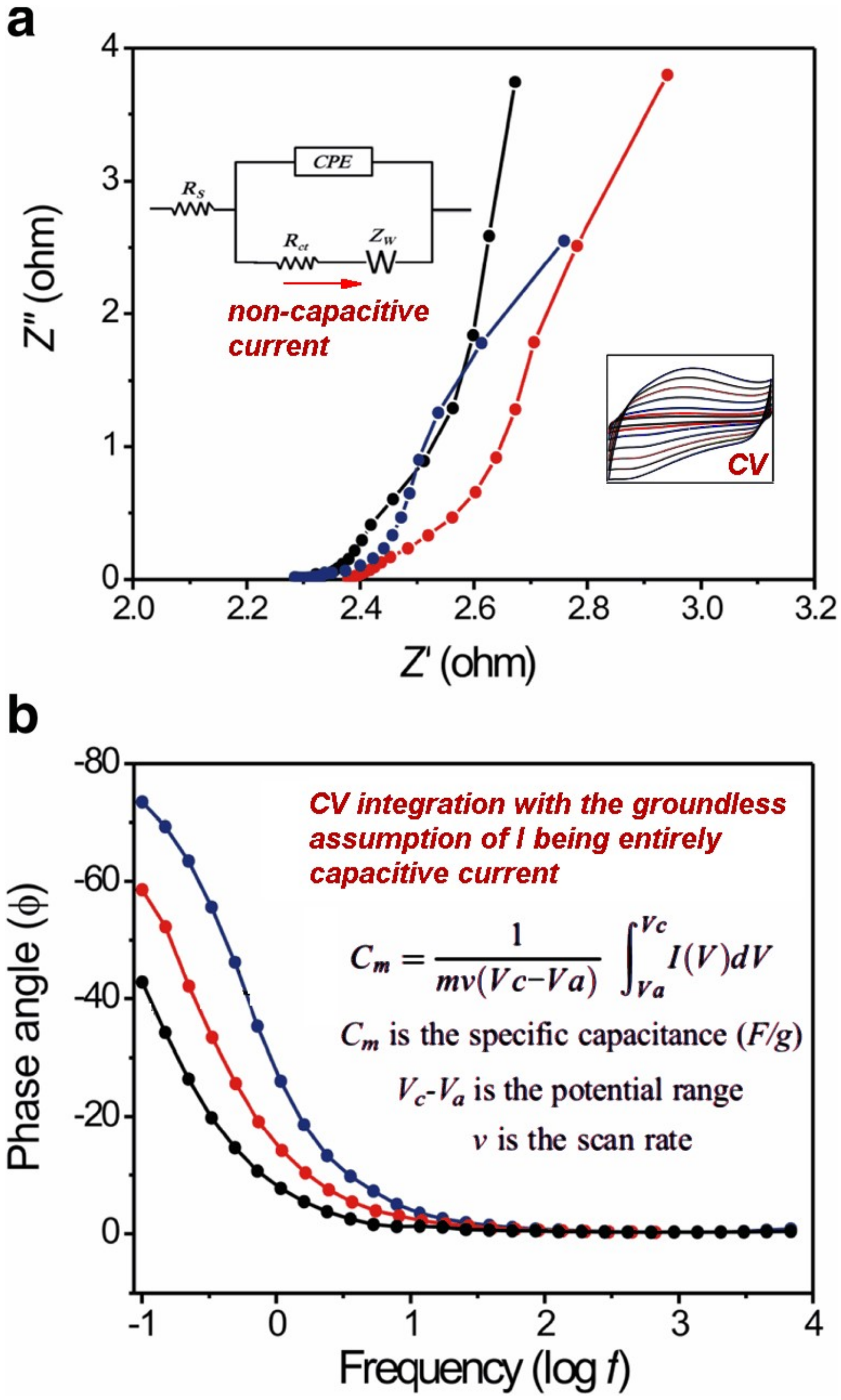

Fig. 2 (a) Nyquist and (b) Bode plots of the supercapacitor investigated in [17]. Inserts show equivalent circuit, characteristic CV profile and the formula used to derive specific capacitance from CV

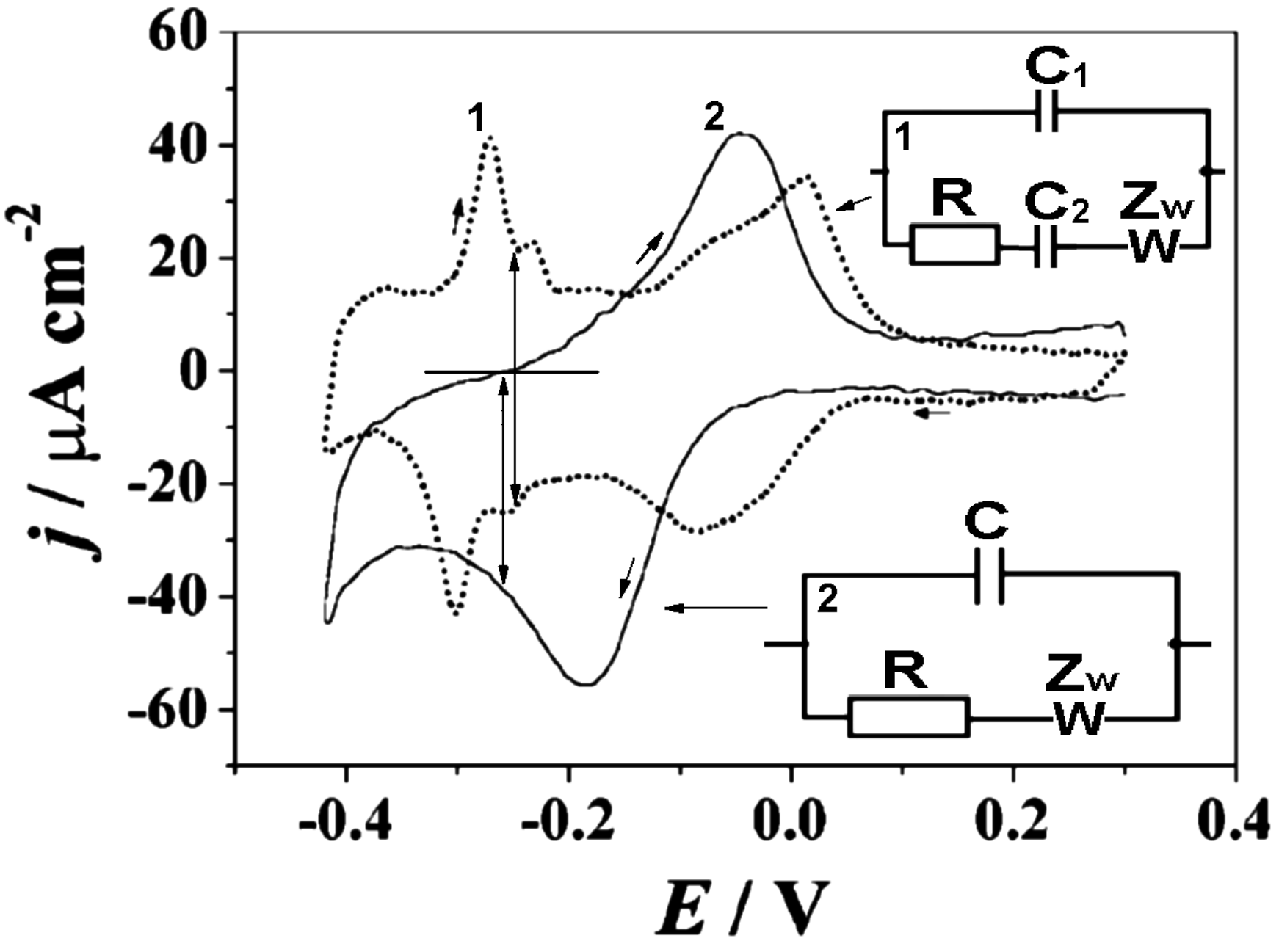

Fig. 3 Cyclic voltammograms of polycrystalline gold electrode in 1mM $Pb(ClO_4)_2$ + 0.1M $HClO_4$ before (1) and after (2) the electrode surface was modified by selenium; the corresponding equivalent electric circuits 1 and 2 of the two systems derived by potentiodynamic electrochemical impedance spectra analysis [19]